\documentclass[conference]{IEEEtran}
\IEEEoverridecommandlockouts
\usepackage{cite}
\usepackage{amsmath,amssymb,amsfonts}
\usepackage{algorithmic}
\usepackage{graphicx}
\usepackage{textcomp}
\usepackage{xcolor}
\usepackage{bbm}
\usepackage{amssymb}
\usepackage{subcaption}
\usepackage{hyperref}

\usepackage{xcolor}

\makeatletter 
\newcommand{\linebreakand}{%
\end{@IEEEauthorhalign}
\hfill\mbox{}\par
\mbox{}\hfill\begin{@IEEEauthorhalign}
}
\makeatother 

\newcommand{\E}[1]{{\mathbbm{E}}\left[ #1 \right]}
\newcommand{\pr}[1]{{\mathbbm{P}}\left[ #1 \right]}
\global\long\def\tr#1{\operatorname{Tr}\left[#1\right]}%
\newcommand{\C}{\mathbb{C}}
\newcommand{\R}{\mathbb{R}}
\newcommand{\ket}[1]{\ensuremath{\left\vert #1 \right\rangle}}
\newcommand{\bra}[1]{\ensuremath{\left\langle #1 \right\vert}}
\newcommand{\ketbra}[2]{\ensuremath{\ket{#1}\!\bra{#2}}}
\DeclareMathOperator*{\argmax}{arg\,max}
\DeclareMathOperator*{\argmin}{arg\,min}

\usepackage{todonotes}

\def\BibTeX{{\rm B\kern-.05em{\sc i\kern-.025em b}\kern-.08em
    T\kern-.1667em\lower.7ex\hbox{E}\kern-.125emX}}
\begin{document}

\title{Vector Field Visualization of Single-Qubit State Tomography \\
}

\author{\IEEEauthorblockN{Adrien Suau}
  \IEEEauthorblockA{\textit{LIRMM, University of Montpellier} \\
    Montpellier, France \\
    \textit{CERFACS},
    Toulouse, France \\
    0000-0002-2412-7298}
  \and
  \IEEEauthorblockN{Marc Vuffray}
  \IEEEauthorblockA{\textit{Theoretical Division} \\
    \textit{Los Alamos National Laboratory}\\
    Los Alamos, NM 87545, USA \\
    0000-0001-7999-9897}
  \and
  \IEEEauthorblockN{Andrey Y. Lokhov}
  \IEEEauthorblockA{\textit{Theoretical Division} \\
    \textit{Los Alamos National Laboratory}\\
    Los Alamos, NM 87545, USA \\
    0000-0003-3269-7263}
  \and
  \linebreakand
  \IEEEauthorblockN{Lukasz Cincio}
  \IEEEauthorblockA{\textit{Theoretical Division} \\
    \textit{Los Alamos National Laboratory}\\
    Los Alamos, NM 87545, USA \\
    0000-0002-6758-4376}
  \and
  \IEEEauthorblockN{Carleton Coffrin}
  \IEEEauthorblockA{\textit{Advanced Network Science Initiative} \\
    \textit{Los Alamos National Laboratory}\\
    Los Alamos, NM 87545, USA \\
    0000-0003-3238-1699}%
}
\maketitle
\begin{abstract}
  As the variety of commercially available quantum computers continues to increase so does the need for tools that can characterize, verify and validate these computers.
  This work explores using quantum state tomography for characterizing the performance of individual qubits and develops a vector field visualization for presentation of the results.
  The proposed protocol is demonstrated in simulation and on quantum computing hardware developed by IBM.
  The results identify qubit performance features that are not reflected in the standard models of this hardware, indicating opportunities to improve the accuracy of these models.
  The proposed qubit evaluation protocol is provided as free open-source software to streamline the task of replicating the process on other quantum computing devices.
\end{abstract}


\section{Introduction}

The emergence of commercial Quantum Computing (QC) hardware has made tools for quantum characterization, verification, and validation (QCVV) \cite{Eisert2020} more important than ever.
Especially in the era of Noisy Intermediate-Scale Quantum (NISQ) \cite{Preskill2018quantumcomputingin} devices, QCVV methods provide the means for QC users to measure and quantify the performance of quantum hardware platforms and enables consistent comparisons across different hardware architectures. 
The scope of QCVV is broad and ranges from testing individual quantum operations (e.g., error rates of one- and two-qubit gates \cite{Wright2019}), verifying small circuits (e.g., Quantum State Tomography \cite{PhysRevA.55.R1561}, Randomized Benchmarking \cite{PhysRevLett.106.180504,PhysRevA.77.012307}, Gate Set Tomography \cite{2009.07301}), to full system-level protocols (e.g., quantum volume estimation \cite{PhysRevA.100.032328}, random quantum circuits \cite{Boixo2018}).
Over the years these QCVV tools have become an invaluable foundation for benchmarking and measuring progress of quantum processors \cite{ibm_qv}, culminating in multiple quantum supremacy demonstrations \cite{Arute2019,PhysRevLett.127.180501}.

In many ways quantum state tomography provides a gold standard for QCVV in that, it can theoretically provide an exact reconstruction of the full quantum state of QC hardware, with sufficient data.  
At its most basic level, quantum state tomography provides a protocol for combining multiple observations to uniquely identify the state of a quantum system (i.e., a density matrix).
On small quantum systems, where data collection and result computations are feasible, quantum state tomography provides a precise measure of how accurately a quantum hardware device can execute a desired quantum computation.
The strength of quantum state tomography for QCVV is that it can provide a comprehensive picture of hardware performance.
The weakness of quantum state tomography is that it often requires a prohibitively large amount of data and the interpretation of the results can be difficult.
Indeed, the objective of many other QCVV protocols (e.g. Randomized Benchmarking \cite{PhysRevLett.106.180504,PhysRevA.77.012307}, Gate Set Tomography \cite{2009.07301}) is to provide trade offs in data collection and result detail resulting in more scalable QCVV alternatives to quantum state tomography.

This work explores the use of quantum state tomography for conducting QCVV on commercially available QC platforms.
We focus on quantum state tomography of single-qubits to minimize the data collection requirements and to develop a protocol that can be executed in parallel for all of the qubits in a given hardware platform.
The core contributions of this work are a {\em Vector Field Visualization} of single-qubit quantum state tomography and an open-source software tool for data collection, state reconstruction and result presentation.
Through experiments on QC hardware available in IBM's Q-Hub, we show that the proposed method can identify qubit performance features that are not easily identified with captured with a single value and provide clear signatures that distinguish qubit performance from both perfect and noisy simulators of this hardware.

This work begins by introducing the foundations of quantum state tomography for a single-qubit in Section \ref{sec:sqt} and reviews the maximum likelihood estimation method \cite{PhysRevA.55.R1561} for reconstructing a quantum state from measurements.
Section \ref{sec:vfvsqt} then proposes the vector field visualization for presenting the single-qubit quantum state tomography results and illustrates how this visualization can be leveraged to provide unique insights into qubit performance.
Section \ref{sec:corruption} investigates how the results from different quantum state tomography algorithms can be combined with the vector field visualization to identify signatures of data corruption.
The details of the open-source software are provided in Section \ref{sec:software} and the paper concludes with a discussion of the usefulness of the proposed protocol and future work in Section \ref{sec:conclusion}.

\section{Single-Qubit State Tomography}
\label{sec:sqt}

The state of a quantum system composed of $n$ qubits is fully described by a $2^n \times 2^n$ Hermitian matrix $\rho$, the so-called density matrix. Density matrices are positive semi-definite, normalized matrices, i.e., $\bra{\phi}\rho\ket{\phi} \geq 0$ for all $\ket{\phi} \in \C^{2^n}$, and $\tr{\rho}=1$.
A measurement outcome of a quantum system is represented by a real random variable, $A$, associated with a Hermitian matrix, customarily also denoted by the same letter.
Measuring $A$ is linked in expectation to the density matrix through the well-known formula  $\E{A \mid \rho} = \tr{\rho A}$. This expression implies that only projections of the density matrix can be observed, as the trace is an inner product over Hermitian matrices.

The task of reconstructing density matrices from repetitive observations of these projections is called Quantum State Tomography (QST). Since the space of density matrices has $4^n -1$ real parameters, exact QST can only be done if one records at least $4^n -1$ different projections. Therefore, general QST remains prohibitive for quantum systems of moderate to large size due to the exponential growth in the number of required observations. However, for single-qubits, QST is tractable and provides a complete description of the quantum system, making it a powerful tool for conducting QCVV of individual qubits in QC hardware.

\subsection{Maximum-Likelihood Quantum State Tomography}
There exists multiple methods (i.e., statistical estimators) that can be used for QST. Each of these estimators comes with its own advantages and disadvantages such as improved reconstruction quality with respect to specific metrics, ease of implementation, or computational benefits.
Some popular choices for QST include linear regression based methods~\cite{Linear_reg_QT_2013}, nuclear norm constrained reconstructions~\cite{nuclear_norm_QT_2010} and the Maximum-Likelihood Estimator (MLE)~\cite{MLE_QT_1997}.
In this work, we adopted the MLE method as it is widely used in practice, leverages fundamental statistical theory principles and can be easily implemented for small quantum systems.
However, a sensitivity study on simulated data suggested that all of these methods produce similar results under the specific data collection settings used in this work.

The Maximum-Likelihood approach for QST consists in finding the density matrix $\rho$ that will maximize the probability of realized measurements. Our observables are measures $k$ described by an ensemble of projectors $\left\{P_k\right\}$ that is the union of all measurement bases or more precisely the Projection-Valued Measures (PVMs) that we consider.
What is recorded, and what serves as an input to the MLE algorithm, are the number of times, $n_k$, that a particular measure $k$ has been observed. The log-likelihood functions is then expressed in terms of our statistics as follows,
\begin{align}
  \ln \pr{\left\{ n_k\right\} \mid \rho} = \sum_{k} n_k \ln \tr{\rho P_{k}}.
\end{align}
The reconstructed density matrix $\rho_{\rm{out}}$ is the output of the following concave maximization problem on the positive semi-definite cone,
\begin{equation}
  \rho_{\rm{out}} = \argmax_{\substack{\rho \succcurlyeq 0 \\[.2em] \tr{\rho} = 1}} \sum_{k} n_k \ln \tr{\rho P_{k}}\label{eq:MLE_QT}
\end{equation}
Due to the limitations of off-the-shelf optimization software, in practice, enforcing the positive semi-definite (PSD) constraint, $\rho \succcurlyeq 0$, may require a specialized optimization algorithm.
One possible approach to solving \eqref{eq:MLE_QT} consists in using local gradient ascent algorithm interleaved with eigendecomposition based projections onto the SDP cone.
However, we will see that for single-qubit QST, the SDP constraint has a convenient simplification.

\subsection{Specializing to Single-Qubit State Tomography}
\label{sec:specializing-single-qubit-tomography}

The Bloch vector representation of quantum states enables us to significant simplify QST for an individual qubit.
In this representation, density matrices are encoded as vectors $\vec{a}\in \R^3$ by $\rho = \frac{1}{2} \left(I + \vec{a} \cdot \vec{\sigma} \right)$ where $\vec{\sigma}$ is the vector of Pauli matrices. The PSD constraint that $\rho$ must satisfy is enforced through the requirement that its corresponding vector lies within the unit sphere, i.e. $\|\vec{a}\| \leq 1$. Projectors are described in a similar fashion by $P_{\vec{u}} = \frac{1}{2} \left(I + \vec{u} \cdot \vec{\sigma} \right)$, where $\|\vec{u}\|=1$. Any PVM is composed of only two projectors $P_{\vec{u}}$ and $P_{-\vec{u}} = I - P_{\vec{u}}$, and therefore can be identified by a single unit vector $\vec{u}$. Since we choose to perform the same number of measurements $N$ in each PVMs, we only need to record the empirical probability $p_{\vec{u}} = n_{\vec{u}} / N$ of measuring the observable associated with $\vec{u}$. The MLE estimator from Eq.~\eqref{eq:MLE_QT} is then simplfied into the following program,
\begin{align}
  \vec{a}_{\rm{out}} = \argmax_{\|\vec{a}\|\leq 1} \sum_{\vec{u}} & p_{\vec{u}} \ln \left(1 + \vec{a}\cdot\vec{u} \right)  +\left(1- p_{\vec{u}}\right) \ln \left(1 - \vec{a}\cdot\vec{u} \right). \label{eq:MLE_QT_single_qbit}
\end{align}
The maximization problem in Eq.~\eqref{eq:MLE_QT_single_qbit} is amenable to standard optimization software for it is a simple concave non-linear problem in $\R^3$ with the cumbersome SDP constraint from Eq.~\eqref{eq:MLE_QT} replaced with a unit sphere constraint.

\subsection{Single-Qubit State Tomography Experiment Design}
\label{sec:experimental_QT}

\begin{figure}[htbp]
  \centerline{\includegraphics[width=0.35\linewidth]{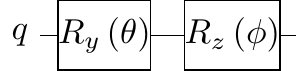}}
  \caption{The state preparation procedure used to initialise a single-qubit quantum state where the elementary rotations are defined by $R_y\left(\theta\right)= \exp\left(- i Y \theta/2\right)$ and $R_z\left(\phi\right) = \left(- i Z \phi /2\right)$.}
  \label{fig:sqp}
\end{figure}

Our main goal is to reconstruct the state of a single qubit programmed to be in the pure state $\rho_{\mathrm{in}} := R_{\theta,\phi} \ketbra{0}{0} R_{\theta,\phi}^\dagger$, where the rotation matrix is implemented through elementary rotations on the $y$ and $z$ axis $R_{\theta,\phi} = R_z(\phi)  R_y(\theta)$ as depicted in Figure~\ref{fig:sqp}.
The programmed density matrix $\rho_{\mathrm{in}}$ is represented on the Bloch sphere with the unit vector $\vec{a}_{\rm{in}}(\theta,\phi) = \left(\sin(\theta) \cos(\phi), \sin(\theta) \sin(\phi),\cos(\theta)\right)$. Using the MLE estimator from Eq.~\eqref{eq:MLE_QT_single_qbit}, we will assess the quality of the state preparation for a specific qubit by comparing $\vec{a}_{\rm{in}}(\theta,\phi)$ with the corresponding $\vec{a}_{\rm{out}}$ for multiple values of $\phi$ and $\theta$.

\begin{figure}[htbp]
  \centerline{\includegraphics[width=\linewidth]{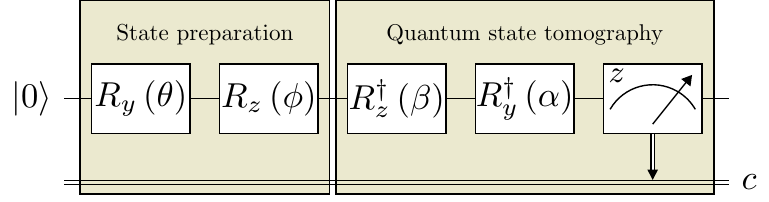}}
  \caption{The state preparation and tomography procedure used to initialise a single-qubit quantum state.}
  \label{fig:sqptm}
\end{figure}

In our experiments, we are limited to measurements in the computational basis, that is to say, to the PVM associated with the vector $(0,0,1)$. To remedy to this issue, we rotate the qubit before measuring it with the matrix $R_{\alpha,\beta}^{\dagger}$ to emulate a PVM on $\vec{u} = \left(\sin(\alpha) \cos(\beta), \sin(\alpha) \sin(\beta),\cos(\alpha)\right)$. The rotation matrix is again implemented with elementary rotations  $R_{\alpha,\beta}^{\dagger} = R_y(- \alpha) R_z(-\beta)$. The complete circuit is shown in Figure \ref{fig:sqptm}. 

\begin{figure}[h]
  \centering
  \includegraphics[width=.47\linewidth]{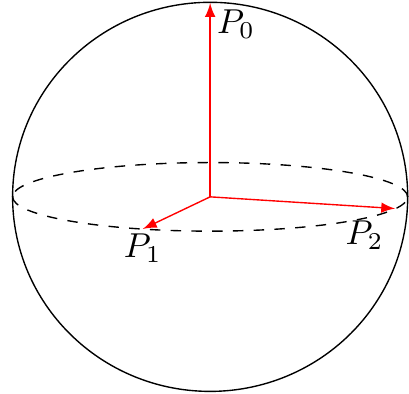}
  \hfill{}
  \includegraphics[width=.48\linewidth]{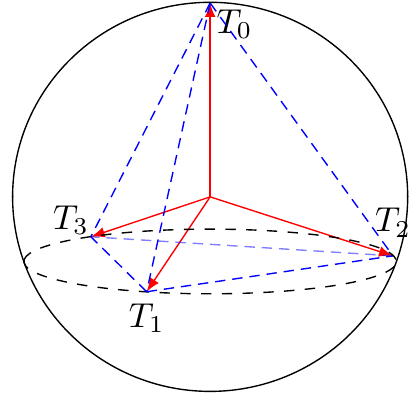}
  \caption{The Bloch sphere representation of PVMs that can be considered for single-qubit state tomography. The Pauli operators (left) provide a minimal set of PVMs while the tetrahedral PVMs used in this work (right) provides data redundancy and improved reconstruction accuracy. In this figure, $P_0 = \ket{0}$, $P_1 = \frac{1}{\sqrt{2}}\left(\ket{0} + \ket{1}\right)$, $P_2 = \frac{1}{\sqrt{2}} \left(\ket{0} + i \ket{1}\right)$, $T_0 = \ket{0}$ and $T_k = \frac{1}{\sqrt{3}}\ket{0} + \sqrt{\frac{2}{3}} e^{i \frac{2}{3}k\pi}\ket{1}$ for $k\in \{1, 2, 3\}$.}%
  \label{fig:bloch_sphere_representation}
\end{figure}

Conducting QST on a single qubit requires measurements from at least 3 different linearly independent PVMs.
One natural choice are the Pauli PVMs along canonical axes (i.e., $(\alpha, \beta) \in \left\{ (0,0), (\pi/2, 0), (\pi/2, 0) \right\}$).
However, having more than the minimum number of PVMs can be beneficial in reducing the statistical error in state reconstruction.
In this work we leverage a tetrahedral set of PVMs using $(\alpha, \beta) \in \left\{ (0,0), (\cos^{-1}(\frac{-1}{3}), 0), (\cos^{-1}(\frac{-1}{3}), \frac{2\pi}{3}), (\cos^{-1}(\frac{-1}{3}), \frac{-2\pi}{3}) \right\}$ to provide a balance of mitigating statistical error and data collection requirements. The Bloch sphere vectors representing both the Pauli and tetrahedral PVMs are depicted in Figure~\ref{fig:bloch_sphere_representation}. 

The experimental procedure for conducting single-qubit QST proposed in this work uses the tetrahedral PVMs with the MLE model from \eqref{eq:MLE_QT_single_qbit} as follows.
For a given state $\vec{a}_{\rm{in}}\left( \theta,\phi \right)$ on the Bloch sphere, four variants of the tomography circuit (Figure \ref{fig:sqptm}) are executed, one for each ($\alpha$, $\beta$) combination in the tetrahedral PVMs.
State measurement statistics are collected for each of these circuits and converted into empirical probabilities $p_{\vec{u}}$ providing the information required for posing the MLE problem \eqref{eq:MLE_QT_single_qbit}.  
The optimal solution to \eqref{eq:MLE_QT_single_qbit}, $\vec{a}_{\rm{out}}$, encodes the most likely density matrix that was implemented by the qubit at the state $\vec{a}_{\rm{in}}\left( \theta,\phi \right)$.

An important subtlety in using a QST workflow in practice is to quantify how the accuracy of the empirical probability impacts the solution quality of model \eqref{eq:MLE_QT_single_qbit}.
On a quantum hardware platform one only has access to a finite number of measurements, $N$ (a.k.a., {\em shots}), to estimate the empirical probabilities $p_{\vec{u}}$.
If the number of measurements is not sufficient, finite sample errors can yield significant artifacts in $\vec{a}_{\rm{out}}$.
We performed a sensitivity study to quantify this statistical error on simulated data. For a given prepared state $\vec{a}_{\rm{in}}\left(\theta, \phi\right)$, we sample $N=20,000$ observations for each of the tetrahedral PVMs, and then reconstruct $\vec{a}_{\rm{out}}$ with our QST MLE algorithm. We repeat this procedure $10^4$ times to accumulate an empirical histogram of the statistical error for each state, $\left(\theta, \phi \right)$, and estimate the $99$-th percentile of the Euclidean distance error $\epsilon$ (i.e., $\pr{\|\vec{a}_{\rm{out}} - \vec{a}_{\rm{in}} \|\leq \epsilon} = 0.99$). We find empirically that $\epsilon \leq 0.02$ for all angles $\left(\theta, \phi \right)$.
Therefore, throughout this work we standardized on $N=20,000$ observations per circuit to ensure that statistical fluctuations will only contribute to an error in the second digit, with high confidence.

\section{Vector Field Visualisation of Single-Qubit State Tomography}
\label{sec:vfvsqt}

To illustrate the potential usefulness of single-qubit QST, this work investigates two questions arising in QCVV: (1) how accurate is single-qubit state preparation; (2) how consistent is the state preparation quality throughout the Bloch sphere. 
The first question amounts to quantifying the difference between the \emph{ideal} quantum state, $\vec{a}_{\rm{in}}(\theta,\phi)$,  and the \emph{reconstructed} quantum state resulting from the QST protocol, $\vec{a}_{\rm{out}}$.
The second question consists of repeating the state tomography protocol for a wide variety of possible state preparations and investigating how the reconstructed states vary. 
Throughout this section we develop the idea of a vector field visualisation for presenting the results of single-qubit QST to visually investigate these two questions. A variety of examples are then used to illustrate the usefulness of the proposed approach.

For a given single-qubit state, $\vec{a}_{\rm{in}}(\theta,\phi)$, the tomography procedure described in Section \ref{sec:sqt} yields a density matrix $\rho_{\rm{out}}$ in the form of a vector in $\vec{a}_{\rm{out}} \in \R^3$.
Many metrics can be computed from $\vec{a}_{\rm{out}}$ such as quantum state purity ($\tr{\rho_{\rm{out}}^2}$) or fidelity ($\tr{\sqrt{\sqrt{\rho_{\rm{in}}}\rho_{\rm{out}}\sqrt{\rho_{\rm{in}}}}}$). 
A natural way to visualize $\vec{a}_{\rm{out}}$ is a point within the Bloch sphere using the relation $\rho = \frac{1}{2} \left(I + \vec{a} \cdot \vec{\sigma} \right)$ from Section \ref{sec:sqt}.
However, this representation is difficult to interpret without an interactive visualization as many points with-in the sphere are co-located in standard orthographic projections, such as Figure \ref{fig:bloch_sphere_representation}.
In particular, it is hard to distinguish if a given point is ``on the front'' or ``on the back'' of the sphere, it is hard to distinguish a pure state, that is represented on the surface of the Bloch sphere, from a mixed state, that is represented strictly within the Bloch sphere.

This work carefully combines three visualization tools to address the challenge of presenting this data.
The first idea is to embed the single qubit states on the two-dimensional plane using established spherical projection methods.
In this work we leverage the \emph{Robinson projection} to place the QST results on a two-dimensional plot.\footnote{Note that any $2$-dimensional projection of a $3$-dimensional sphere is bound to imperfectly represent some of the features.
The Robinson projection is a compromise and is neither angle-preserving nor equal-area, but a balance designed to reduce the overall distortion.}
The second idea is to leverage a heatmap on this $2$-dimensional representation to visualize a key metric of interest, such as the reconstructed state's purity or fidelity.
Throughout this work the color of the plot is used to present the state's purity.
The third idea is to use arrows to illustrate the locations difference between the \emph{ideal} and \emph{reconstructed} states, with each arrow starting at the \emph{ideal} state and ending at the corresponding \emph{reconstructed} state, projected onto the surface of the Bloch sphere.
This communicates the rotational error that occurs in the \emph{reconstructed} state. 
Overall, we call this representation of single-qubit QST the Vector Field Visualisation (VFV).

\subsection{Vector Field Visualisation Examples}

\begin{figure}[tbp]
     \centering
     \begin{subfigure}{\linewidth}
         \centering
         \includegraphics[width=\linewidth]{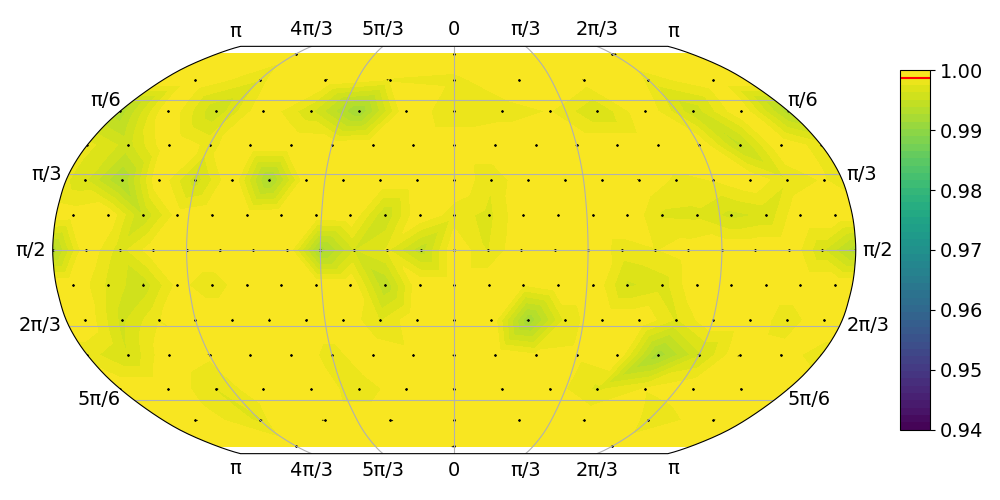}
         \caption{Data obtained on an ideal simulator.}%
         \label{fig:lagos-ideal-simulator-purity}
     \end{subfigure}
     \begin{subfigure}{\linewidth}
         \centering
         \includegraphics[width=\linewidth]{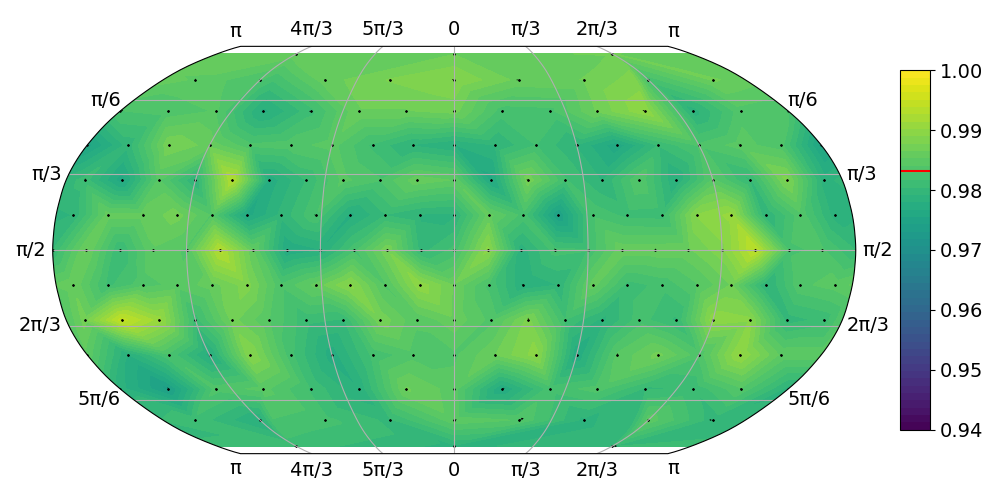}
         \caption{Data obtained on a noisy simulator configured with the calibration data of \texttt{ibm\_lagos} at the time of hardware data collection.}%
         \label{fig:lagos-noisy-simulator-purity}
     \end{subfigure}
     \begin{subfigure}{\linewidth}
         \centering
         \includegraphics[width=\linewidth]{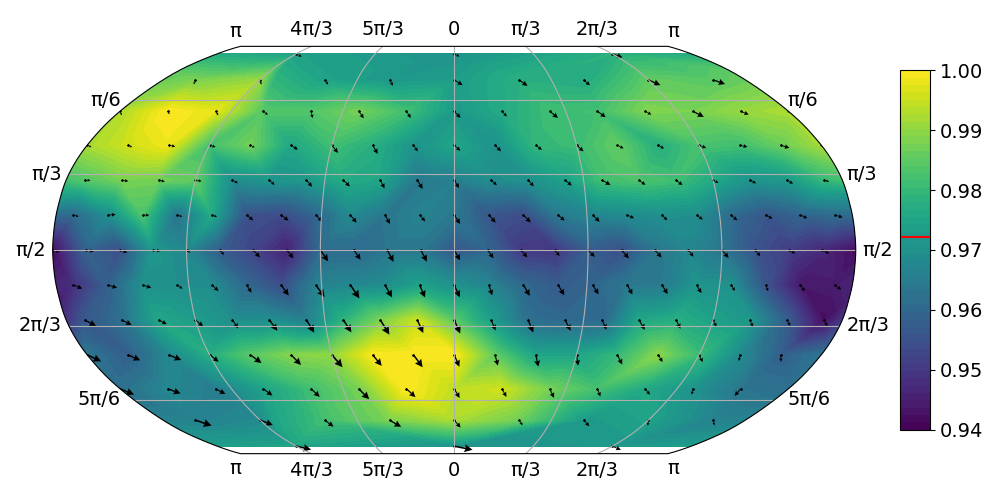}
         \caption{Data obtained on \texttt{ibm\_lagos}.}%
         \label{fig:lagos-noisy-chip-purity}
     \end{subfigure}
  \caption{Vector Field Visualisations of the reconstructed states using the single-qubit state tomography protocol from Section~\ref{sec:sqt} using $20,000$ shots on qubit $1$ of \texttt{ibm\_lagos}. The heatmap indicates the purity of the reconstructed quantum state and the horizontal red line in the color scale indicates the average purity of the states.}%
  \label{fig:vfvsqt}
\end{figure}

To verify the correctness of the proposed VFV approach, and to illustrate its usefulness, we begin with a series of QST studies of $200$ quantum states spaced approximately equidistantly around the Bloch sphere using the \texttt{ibm\_lagos} computer.
The results are presented in Figure \ref{fig:vfvsqt}.
As a validation exercise, we begin by performing the full QST protocol on an ideal quantum simulator in Figure \ref{fig:lagos-ideal-simulator-purity}.
This procedure yields reconstructed states with very high quality  (purity $0.999$, on average) with no rotational errors, confirming the correctness of the workflow's implementation.
The slight variability in these results illustrates the small errors that occur due to finite samples.

The second experiment, Figure \ref{fig:lagos-noisy-simulator-purity}, consists of performing QST on data from a simulator of a noisy quantum computer.
IBM's Q-Hub tracks various performance properties of their qubits (e.g., gate and measurement errors) and provides an interface to perform noisy simulations using these properties.
As expected, this procedure yields less consistent reconstructed states (purity $0.983$, on average) with no rotational errors.
Most notably, the simulated noisy qubit performance shows no rotational errors and is very homogeneous regardless of the quantum state that is being inspected.

The third and most interesting experiment, Figure \ref{fig:lagos-noisy-chip-purity}, consists of performing the state tomography protocol with real data from the \texttt{ibm\_lagos} computer.
Three interesting observations can be made when comparing this result to the two simulations: (1) This is the first dataset where the vector field arrows are visible, indicating a type of rotational error that is not captured the simulations; (2) The purity of the reconstructed states is more heterogeneous than the simulators; (3) the purity of the reconstructed states is similar in average to the noisy simulator ($0.983$ compared to $0.972$) but has a notably higher standard deviation ($0.004$ compared to $0.013$).
All of these observations indicate the potential for a state-dependent error model to better capture this qubit's performance.
Overall, the value of VFV approach is highlighted by the distinct motifs that are clear in each of these figures and may provide inspiration for new qubit performance measures and noise mitigation schemes.

\subsection{Visualisation of State Degradation}

\begin{figure}[tbp]
  \centerline{\includegraphics[width=0.5\linewidth]{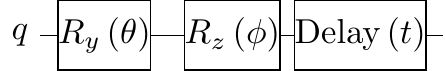}}
  \caption{Quantum state-preparation circuit used to visualise the effect of state degradation over time. The original protocol is obtained by setting the delay time $t = 0$. Increasing $t$ shows the errors that arise in idle open-quantum systems. On IBM Quantum chips, $t$ is given as a multiple of a specific time $\rm{dt}$. On \texttt{ibm\_lagos}, $\rm{dt} = \frac{2}{9}\rm{ns}$.}
  \label{fig:sqp-delay}
  \vspace{-0.5cm}
\end{figure}

One of the primary uses of data presentation tools like the VFV is to examine features of qubit performance that are not captured by current QC simulators.
This can provide inspiration for which features could make simulators more faithful proxies of real-world QC hardware.
To illustrate the point this section explores the impacts of adding a \texttt{delay(t)} instruction into the state preparation protocol, as shown in Figure \ref{fig:sqp-delay}.
In this circuit the state preparation procedure used in the original protocol is now followed by a delay during which the qubit experiences the effects of an open-quantum system before the QST measurements are performed.
Note that this delayed QST experiment would produce identical results in the case of {\em closed-system} simulations that are considered in Figure \ref{fig:vfvsqt}.
This type of experiment is only interesting on QC hardware or simulations of open-quantum systems \cite{2011.14046,JOHANSSON20121760}.

\begin{figure}[htbp]
  \centering
  \begin{subfigure}{\linewidth}
    \centering
    \includegraphics[width=\linewidth]{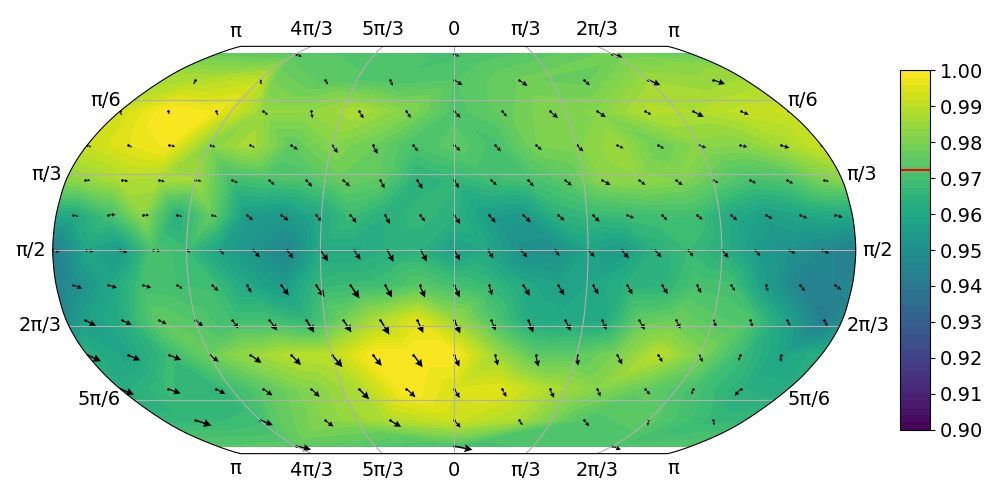}
    \caption{Data obtained on \texttt{ibm\_lagos} with a delay of $t = 0\ \mathrm{dt}$.}%
    \label{fig:lagos-delay-0}
  \end{subfigure}
  \begin{subfigure}{\linewidth}
    \centering
    \includegraphics[width=\linewidth]{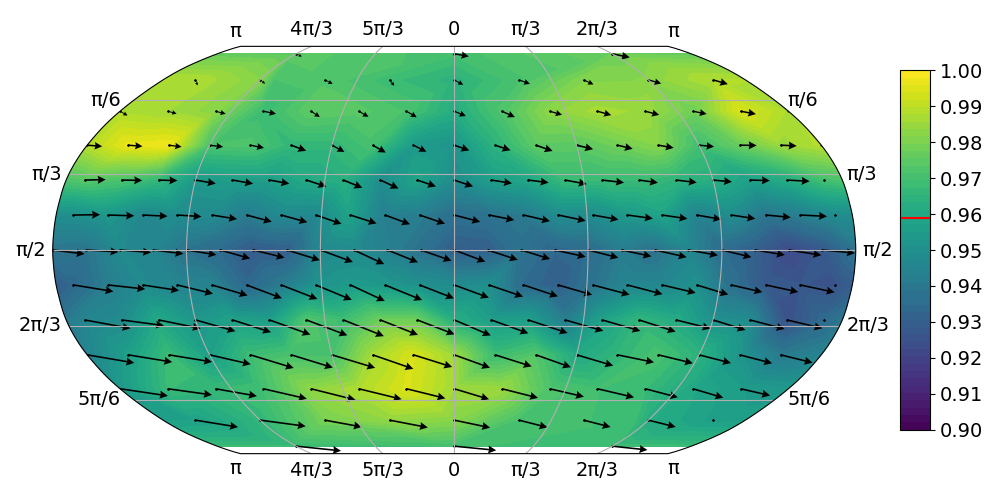}
    \caption{Data obtained on \texttt{ibm\_lagos} with a delay of $t = 800\ \mathrm{dt}$.}%
    \label{fig:lagos-delay-800}
  \end{subfigure}
  \begin{subfigure}{\linewidth}
    \centering
    \includegraphics[width=\linewidth]{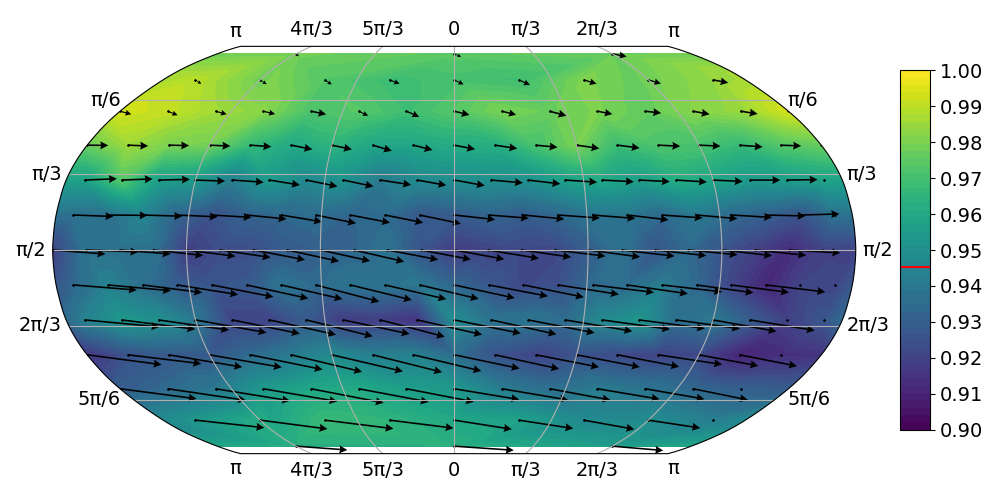}
    \caption{Data obtained on \texttt{ibm\_lagos} with a delay of $t = 1600\ \mathrm{dt}$.}%
    \label{fig:lagos-delay-1600}
  \end{subfigure}
  \caption{Vector Field Visualisations of the reconstructed states of the state preparation with delay circut using $20,000$ shots on qubit $1$ of \texttt{ibm\_lagos} using three different delay values. The heatmap indicates the purity of the reconstructed quantum state and the horizontal red line in the color scale indicates the average purity of the states.}%
  \label{fig:vfvsqt-delay}
\end{figure}

Results of the time delay experiments are shown in Figure \ref{fig:vfvsqt-delay}. 
Two notable observations can be made from these results.
First we can see that the average of the state's purity degrades steadily, starting with an average value of $0.972$ and degrading to $0.959$ and then $0.945$.
The second observation is that the rotational error of the state is also increasing steadily and non-uniformly with time; notice how the rotational bias at the top of the VFV is very different from the bottom.
It is also suprizing to see a systematic rotational shift from left to right that appears to be state-dependent.
Although this study is only a proof-of-principle, it highlights the potential usefulness of VFV for designing models of open-quantum systems and provides some intuition for what decoherence {\em looks like} on this QC hardware.

\section{Signatures of Single-Qubit Data Corruption}
\label{sec:corruption}

In this section we explore how different QST reconstruction methods can be combined to identify additional issues with the performance of individual qubits.
In particular, we compare reconstructed states using the MLE method with those produced by the Linear Regression (LR) method~\cite{Linear_reg_QT_2013}. In short, LR QST identifies a density matrix that minimizes the difference between the observed and predicted measurement probabilities. For a single qubit system, the LR method solves the following optimization task on the Bloch sphere,
\begin{align}
  \vec{a}^{\rm{LR}}_{\rm{out}} = \argmin_{\|\vec{a}\|\leq 1} \sum_{\vec{u}} \left(1+ \vec{u} \cdot \vec{a} - 2p_{\vec{u}}\right)^2. \label{eq:LR_QT_single_qbit}
\end{align}
Using the statistical error analysis from Subsection~\ref{sec:experimental_QT}, we find that the $99$-th percentile of the Euclidean distance error of LR is less than $0.02$ for 20,000 observations, which is comparable to the MLE method.

\begin{figure}[tbp]
  \centering
  \begin{subfigure}{\linewidth}
    \centering
    \includegraphics[width=\linewidth]{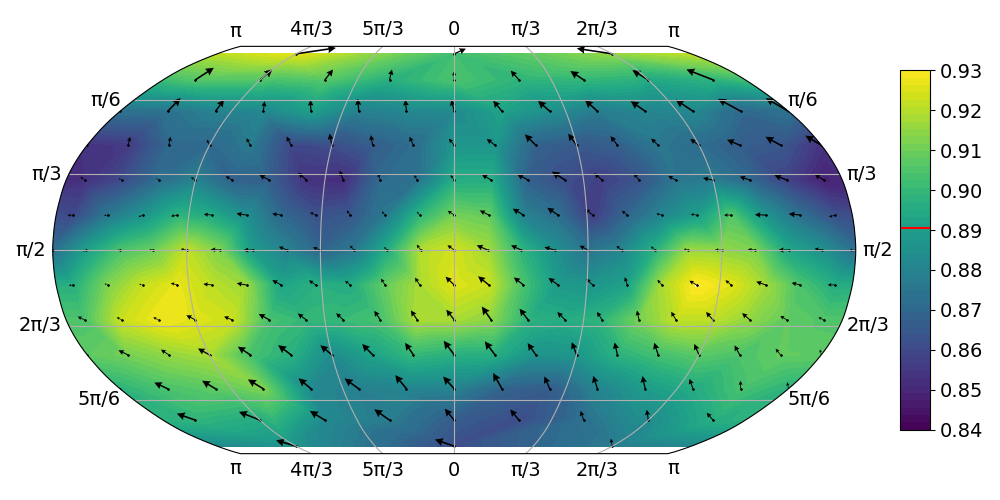}
    \caption{Data obtained on \texttt{ibmq\_belem} and reconstructed with MLE.}%
    \label{fig:belem-spots-mle}
  \end{subfigure}
  \begin{subfigure}{\linewidth}
    \centering
    \includegraphics[width=\linewidth]{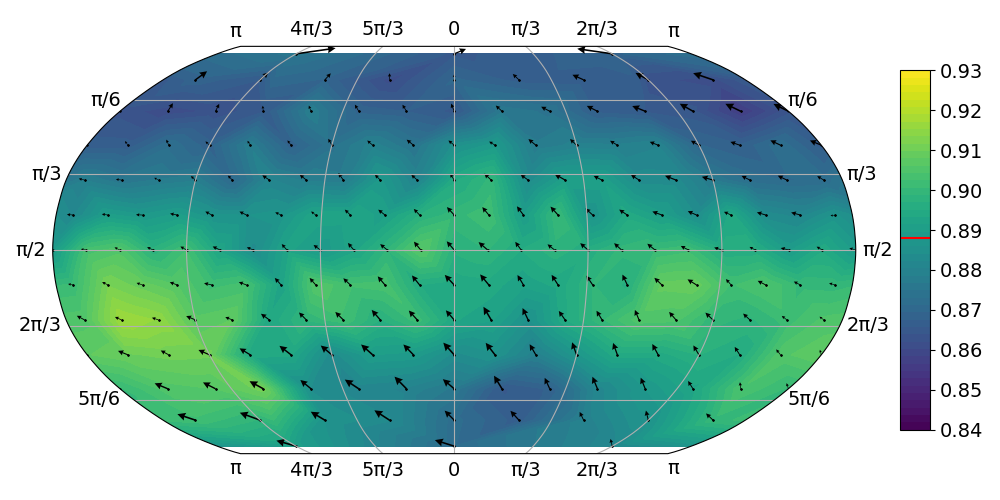}
    \caption{Data obtained on \texttt{ibmq\_belem} and reconstructed with LR.}%
    \label{fig:belem-spots-lssr}
  \end{subfigure}
    \begin{subfigure}{\linewidth}
    \centering
    \includegraphics[width=\linewidth]{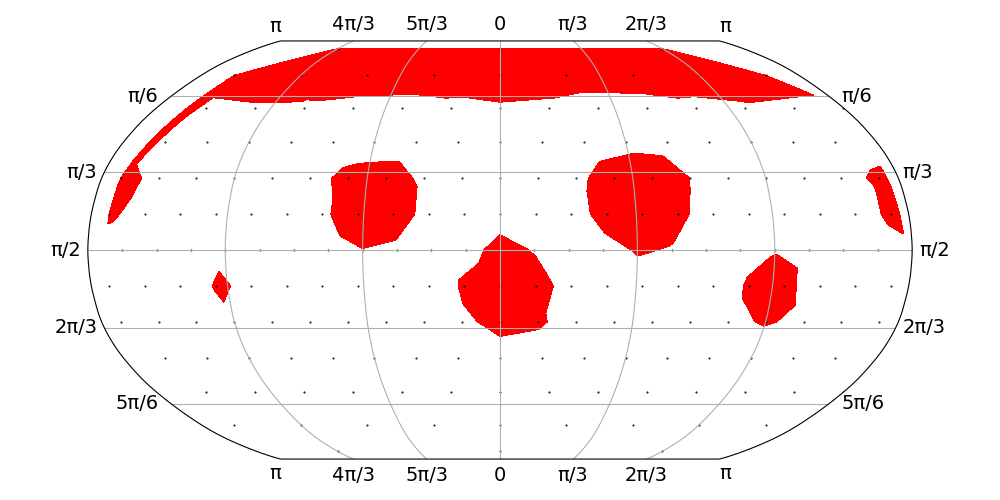}
    \caption{Reconstructed quantum states where the purity difference between MLE and LR reconstruction methods is above the $0.02$ threshold.}%
    \label{fig:belem-spots-difference}
  \end{subfigure}
  \caption{A comparison of the quantum state reconstruction methods MLE and LR on qubit $3$ of \texttt{ibmq\_belem}, where the reconstruction methods do not entirely agree. Each post-processing method is given the same raw data from \texttt{ibmq\_belem}.}%
  \label{fig:mle-lssr-spots-difference}
\end{figure}

The key insight of this section is that the difference in purity between the reconstructed states of MLE and LR should be very small, but in practice is it not always the case.
Specifically, when using identical input statistics, the $99$-th percentile of the Euclidean distance between the state estimates of MLE and LR is also less than $0.02$ with 20,000 observations  (i.e., $\pr{\left|\|\vec{a}^{\rm{LR}}_{\rm{out}}\| - \|\vec{a}^{\rm{MLE}}_{\rm{out}} \|\right|\leq 0.02} \geq 0.99$). 
However, in practice, we observe that some qubits produce data where the differences in state reconstructions by MLE and LR cannot be reasonably explained by statistical fluctuations, as shown in Figure~\ref{fig:mle-lssr-spots-difference}.
This suggests that the the measured statistics of these qubits are corrupted after the state preparation occurs.
It is likely that the elementary rotations $R_y(- \alpha)$ and $R_z(-\beta)$ used for changing measurement basis introduce \emph{state-dependant} errors that are incompatible with the QST models considered in this work.
Nevertheless, the analysis in Figure~\ref{fig:mle-lssr-spots-difference} indicates that combining multiple tomography methods can be a useful and effective tool for identifying signatures of data corruption.

\section{Open-Source Software Implementation}\label{sec:software}

In the last couple of years, IBM's Qiskit python package has emerged as a de facto standard for gate-based QC and can be used to access a wide variety of hardware platforms, even beyond those provided by IBM. As such, the libraries developed in this work leverage Qiskit for data collection on QC hardware. Even though Qiskit is the only supported framework for the moment, particular attention has been given to enable extending to other frameworks in the future.

The single-qubit QST protocol presented in this work is organized into two packages: \texttt{sqt} and \texttt{sqmap}. 
The first package, \texttt{sqt} (short for Single-Qubit Tomography), implements all the functions and interfaces to perform a single-qubit QST efficiently. 
In particular, it provides various single-qubit tomography basis and quantum state reconstruction methods. 
\texttt{sqt} also provides ways to parallelize a single-qubit experiment over all the available qubits in QC hardware for a quick assessment of the qubit performance across a full-chip. 
The second package, \texttt{sqmap} (short for Single-Qubit Map), provides various ways of visualising single-qubit tomography results obtained with \texttt{sqt}, such as the VFV figures presented in this work.

All of the QST procedures and plotting facilities used in this paper are available as open-source at \url{https://github.com/nelimee/sqt} and \url{https://github.com/nelimee/sqmap} and can be used freely by anyone to benchmark their QC platform. 
Installation of these packages can be done with the official Python package manager \texttt{pip}.


\section{Conclusion}
\label{sec:conclusion}

As the variety of quantum computing platforms continues to increase so does the need of tools to inspect their performance characteristics.
In this work we have demonstrated that quantum state tomography of individual qubits is a viable approach for inspecting qubit performance, although similar procedures are unlikely to scale to much larger systems due to the notable data collection requirements.
Through the careful design of data collection, state reconstruction and result visualization, this work illustrates that the proposed QCVV procedures can highlight elusive patterns in qubit performance that are difficult to capture with simpler metrics.
The results indicate that there is room for improvement on the simulation models that are currently used and highlight the importance of modeling open-quantum system effects at medium time scales. 
A careful comparison different tomography methods also indicates that more general models for quantum state tomography should be considered to better capture the exotic effects that can be observed is current QC platforms.
We hope that the proposed QCVV procedure and vector field visualization will be a valuable tool for the quantum computing community in evaluating qubit performance and have provided the implementation as open-source to support that aim.

\section*{Acknowledgment}

This work was partly supported by the U.S. DOE through a quantum computing program sponsored by the Los Alamos National Laboratory (LANL) Information Science \& Technology Institute and the Laboratory Directed Research and Development (LDRD) program of LANL under project numbers 20210116DR, 20210114ER and the Center for Non Linear Studies. This research was also partly supported by the U.S. Department of Energy (DOE), Office of Science, Office of Advanced Scientific Computing Research, under the Accelerated Research in Quantum Computing (ARQC) program.
This research used quantum computing resources provided by the LANL Institutional Computing Program, which is supported by the U.S. Department of Energy National Nuclear Security Administration under Contract No. 89233218CNA000001.
Adrien Suau also thanks TotalEnergies for their general support of this work. Adrien Suau also thank Gabriel Staffelbach and Aida Todri-Sanial for interesting and lively discussions around the paper.

\bibliographystyle{IEEEtran}
\bibliography{main}

LA-UR-22-23640

\end{document}